\documentclass[conference]{IEEEtran}
\IEEEoverridecommandlockouts
\usepackage{cite}
\usepackage{amsmath,amssymb,amsfonts}
\usepackage{algorithmic}
\usepackage{graphicx}
\usepackage{textcomp}
\usepackage{xcolor}
\usepackage{url}
\usepackage{hyperref}
\hypersetup{
    colorlinks=true,
    linkcolor=black,
    citecolor=blue,
    urlcolor=blue
}
\def\BibTeX{{\rm B\kern-.05em{\sc i\kern-.025em b}\kern-.08em
    T\kern-.1667em\lower.7ex\hbox{E}\kern-.125emX}}
\begin{document}

\title{ThinkTank: A Framework for Generalizing Domain-Specific AI Agent Systems into Universal Collaborative Intelligence Platforms}

\author{
\IEEEauthorblockN{
Praneet Sai Madhu Surabhi\IEEEauthorrefmark{1} \hspace{1.5cm} Dheeraj Reddy Mudireddy\IEEEauthorrefmark{2}
}\\
\IEEEauthorblockA{
\IEEEauthorrefmark{1}\IEEEauthorrefmark{2}Department of Computer Science, Texas A\&M Institute of Data Science, Texas A\&M University, College Station, Texas\\
\IEEEauthorrefmark{1}psurabhi@tamu.edu,
\IEEEauthorrefmark{2}dheeraj.reddy@tamu.edu
}
\\[1.2em]
\IEEEauthorblockN{Jian Tao}\\
\IEEEauthorblockA{
College of Performance, Visualization \& Fine Arts,\\
Department of Electrical \& Computer Engineering,\\
Department of Multidisciplinary Engineering,\\
Department of Nuclear Engineering,\\
Texas A\&M Institute of Data Science,\\
Texas A\&M University, College Station, Texas\\
jtao@tamu.edu
}
}

\maketitle

\begin{abstract}
This paper presents ThinkTank, a comprehensive and scalable framework designed to transform specialized AI agent systems into versatile collaborative intelligence platforms capable of supporting complex problem-solving across diverse domains. ThinkTank systematically generalizes agent roles, meeting structures, and knowledge integration mechanisms by adapting proven scientific collaboration methodologies. Through role abstraction, generalization of meeting types for iterative collaboration, and the integration of Retrieval-Augmented Generation with advanced knowledge storage, the framework facilitates expertise creation and robust knowledge sharing. ThinkTank enables organizations to leverage collaborative AI for knowledge-intensive tasks while ensuring data privacy and security through local deployment, utilizing frameworks like Ollama with models such as Llama3.1. The ThinkTank framework is designed to deliver significant advantages in cost-effectiveness, data security, scalability, and competitive positioning compared to cloud-based alternatives, establishing it as a universal platform for AI-driven collaborative problem-solving.
The ThinkTank code is available at \url{https://github.com/taugroup/ThinkTank}.
\end{abstract}


\section{Introduction}
The proliferation of specialized AI agent systems in scientific research has underscored the effectiveness of multi-agent collaboration for tackling complex problem-solving tasks. Recent advancements in Large Language Models (LLMs) have catalyzed the development of sophisticated multi-agent systems (MAS) capable of coordinating and collectively solving intricate problems at scale. This marks a significant transition from isolated, task-specific models to more dynamic, collaboration-centric approaches. Research has demonstrated that autonomous agents can successfully emulate scientific collaboration methodologies to generate evidence-driven hypotheses. Frameworks such as SciAgents \cite{ghafarollahi2024sciagents} and Virtual Lab \cite{Swanson2024.11.11.623004} have showcased the ability to autonomously generate and evaluate promising research hypotheses across various fields through enhanced human-AI collaboration. However, a persistent limitation of these systems is their typical confinement to their original domains, which curtails their broader applicability and overall organizational value. This paper directly addresses the critical challenge of generalizing domain-specific AI agent architectures to forge universal collaborative intelligence platforms. 

ThinkTank is a novel framework designed for the systematic generalization of agent roles, collaboration structures, and knowledge integration mechanisms. Our contributions in this paper are fourfold: (1) we present a comprehensive framework for generalizing scientific collaboration agent roles to universal problem-solving contexts, enabling broader applicability; (2) we detail the adaptation of meeting structures, traditionally used in human collaboration, for diverse productivity applications within the AI agent framework; (3) we provide a technical architecture for dynamic agent creation, integrating Retrieval-Augmented Generation(RAG) to enhance knowledge grounding and contextual relevance; and (4) we support local deployment strategies to bolster privacy and security. By adapting proven scientific collaboration methodologies and leveraging local deployment through frameworks like Ollama with models such as Llama3.1, ThinkTank empowers users to harness collaborative AI for knowledge-intensive tasks while maintaining data sovereignty and cost-effectiveness.

\section{related work}
Multi-agent systems have demonstrated effectiveness across various domains, with particular success in scientific research collaboration \cite{gottweis2025towards} \cite{Shu2024} \cite{tran2025multi}. Recent advances in large language models have enabled more sophisticated agent interactions and knowledge integration capabilities through frameworks that leverage multiple LLM agents to coordinate and solve complex tasks collectively. The AWS research team has extensively worked on building and evaluating multi-agent collaboration (MAC) frameworks, demonstrating that combining the reasoning power of multiple intelligent specialized agents represents a powerful approach to tackle intricate, multistep workflows \cite{Shu2024}.

RAG systems have shown promise for enhancing AI agent capabilities through external knowledge integration. The MAIN-RAG framework introduces multi-agent filtering mechanisms that leverage multiple LLM agents to collaboratively filter and score retrieved documents, demonstrating 2-11\% improvement in answer accuracy while reducing irrelevant retrieved documents \cite{chang2024main}. Local deployment frameworks have emerged as viable alternatives to cloud-based solutions, addressing privacy and cost concerns while maintaining sophisticated AI capabilities \cite{huang2025onpremises}. The integration of these technologies with multi-agent collaboration represents an underexplored area with significant potential for organizational productivity enhancement.

Frameworks like SciAgents \cite{ghafarollahi2024sciagents} and Virtual Lab \cite{Swanson2024.11.11.623004} exemplify the transformative potential of AI-driven scientific discoveries, demonstrating the power of collective intelligence.  SciAgents combines ontological knowledge graphs, LLMs, and multi-agent systems to autonomously generate and refine hypotheses across disciplines, uncovering hidden scientific insights. Virtual Lab, on the other hand, structures a team of specialized LLM agents guided by a principal investigator agent and human feedback to solve complex interdisciplinary problems like nanobody design. While both frameworks illustrate significant strides in domain-specific AI research collaboration, their design and deployment remain tailored to specific scientific workflows. Drawing inspirations from the collaborative AI methodologies demonstrated in the Virtual Lab project, ThinkTank aims to generalize and extend these paradigms by offering a domain-agnostic, modular platform that translates such specialized strategies into a universally applicable agentic collaboration infrastructure.

Privacy-preserving mechanisms for LLMs have become increasingly important, with comprehensive surveys exploring technical foundations including differential privacy, federated learning, cryptographic protocols, and trusted execution environments \cite{zhao2024privacy}. Research demonstrates that deploying closed-source LLMs within user-controlled infrastructure enhances data privacy and mitigates misuse risks through on-premises deployment that ensures model confidentiality while offering privacy-preserving customization \cite{huang2025onpremises}. Local LLM deployment provides unprecedented levels of data protection by keeping sensitive information within organizational control, which is particularly crucial for industries dealing with confidential information such as healthcare, finance, and legal services.

\section{Methodology}
Scientific collaboration, the cooperative effort of researchers, institutions, and communities leveraging diverse expertise to tackle complex problems and drive innovation, serves as a foundational inspiration for the ThinkTank framework \cite{tran2025multi}. This mode of knowledge production is increasingly dominant, particularly as modern research problems demand interdisciplinary approaches and the integration of varied skill sets \cite{schmidgall2025agentrxiv}. The core tenets of scientific collaboration: enhancing research efficiency, fostering knowledge creation, and sharing resources, parallel the objectives of multi-agent AI systems designed for complex problem-solving. The ThinkTank framework draws inspiration from established scientific collaboration practices to generalize agent interactions. Key aspects of human scientific collaboration that inform ThinkTank's design include:

\subsubsection{Interdisciplinary Integration} Scientific progress increasingly relies on bridging disparate fields of knowledge \cite{schmidgall2025agentrxiv}, \cite{tran2025multi}. Similarly, ThinkTank aims to enable AI agents with specialized "knowledge" from different domains to collaborate effectively. This mirrors how human researchers from various disciplines contribute unique perspectives to a common problem.

\subsubsection{Structured Interaction} Effective scientific collaborations often involve defined roles, regular meetings, and clear communication protocols. ThinkTank adapts these concepts by proposing generalized agent roles and standardized meeting structures for AI agents, facilitating coordinated action and information exchange.

\subsubsection{Knowledge Exchange and Co-creation} A central theme in scientific collaboration is the mutual learning and co-creation of knowledge that occurs when individuals with different backgrounds work together \cite{ghafarollahi2024sciagents}, \cite{gottweis2025towards}, \cite{lu2024aiscientist}. The concept of "co-creation," where stakeholders collaboratively develop solutions, is increasingly recognized in research methodologies and can be supported by AI. Each agent in ThinkTank has a memory module that aims to provide agents with access to and the ability to integrate and remember diverse knowledge sources, emulating the rich information environment of collaborative human research.

\subsubsection{Human-AI Collaboration Models} The evolution of AI in science has led to new paradigms in which AI can act as a tool, an assistant, or even a collaborative partner in research \cite{han2024llm}. Future work envisions ThinkTank to become a hybrid collaborative environment where AI agents not only interact with each other but also engage dynamically with human users, fostering a more integrated and adaptive problem-solving ecosystem. The goal is to move towards systems capable of more autonomous, long-term scientific discovery and reasoning, which inherently requires robust collaboration mechanisms.

By abstracting the successful principles observed in human scientific collaboration, ThinkTank seeks to create a versatile platform where AI agents can synergistically combine their capabilities. This approach aims to overcome the limitations of siloed, domain-specific AI systems and foster a more universal and adaptable form of collaborative AI. The challenges in human scientific collaboration, such as ensuring effective communication and integrating diverse perspectives, also offer valuable lessons for designing resilient and effective multi-agent AI systems.

\section{Implementation}
\subsection{System Architecture Design}
\begin{figure*}[t]
    \centering
    \includegraphics[width=1.0\textwidth]{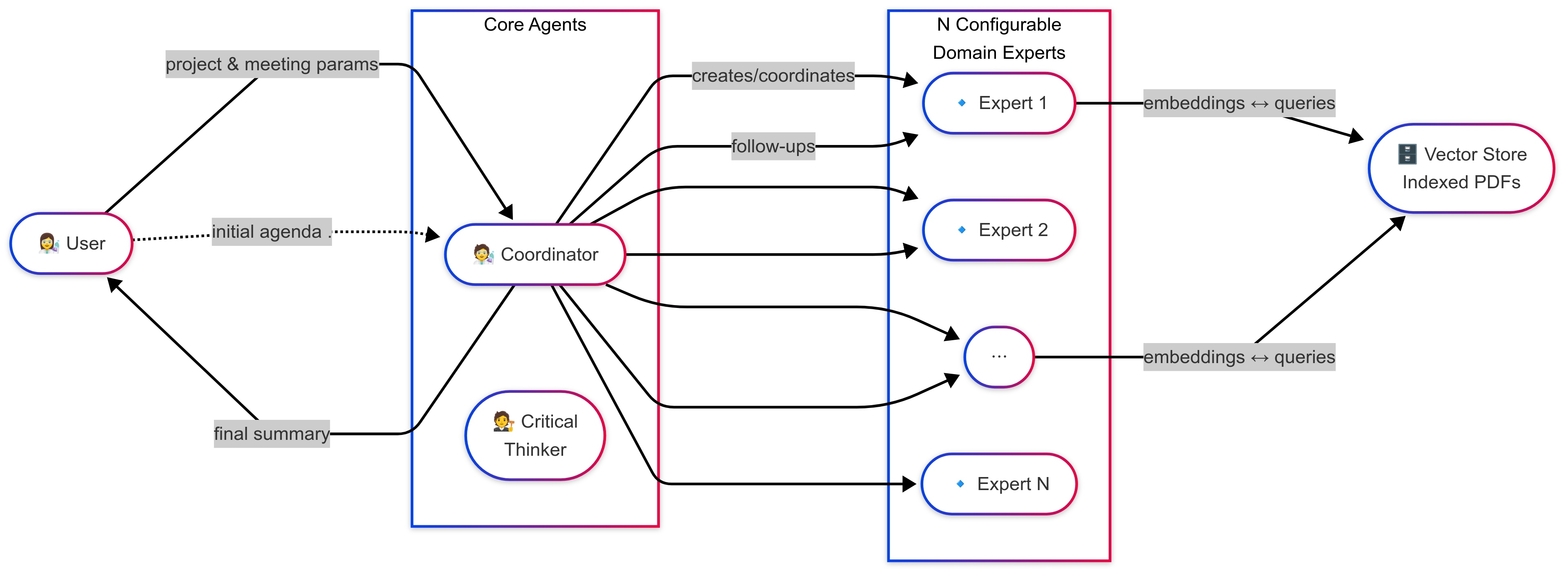}
    \caption{The diagram illustrates the interactions among key components of ThinkTank: Core Agents (Coordinator and Critical Thinker), N Configurable Domain Experts, and a Vector Store of Indexed PDFs. The User initiates the process by providing project and meeting parameters, along with the initial agenda, to the Coordinator. The Coordinator manages the workflow by creating tasks and coordinating with multiple Domain Experts, issuing follow-ups as needed. Each Domain Expert can independently query the Vector Store using embeddings to retrieve relevant information from indexed documents. The Critical Thinker agent supports the Coordinator by providing additional analysis or critique. Upon completion, the Coordinator synthesizes the experts’ input and delivers a final summary back to the User.}
    \label{fig:arc}
    \vspace{-3mm}
\end{figure*}

The ThinkTank system is built upon an open-source Python framework Agno~\cite{agno}, a robust and modular architecture designed to support dynamic multi-agent collaboration. This architecture prioritizes scalability, allowing for horizontal scaling through the addition of more agents and vertical scaling via the expansion of computational resources, thereby addressing common scalability challenges in multi-agent systems where the number of potential interactions can grow exponentially. The architecture integrates several core components, drawing from established frameworks for multi-agent systems and compound AI systems, which often involve LLMs integrated into an expansive software infrastructure with components like databases and tools.
The core components of the ThinkTank architecture include:
\subsubsection{Agent Orchestration Layer} This foundational layer is responsible for managing the entire life cycle of the AI agents. When a user initiates a project, this system parses the project description, objectives, expert configuration and any specified domain requirements. Based on this analysis, it instantiates the necessary specialist Domain Expert Agents (see Section IV.B), along with the Generalized Coordinator and Critical Thinker agents, equipping them with relevant domain-specific capabilities. This automated approach to expertise identification and agent instantiation directly addresses the challenge of optimizing task allocation. Beyond creation, the Orchestration Layer handles inter-agent communication protocols, ensuring seamless information exchange, and manages resource allocation, efficiently distributing computational loads and tasks among the active agents. Effective orchestration, coordinating data and instructions, is crucial for maintaining coherence and efficiency in complex collaborative tasks, with decentralized decision-making approaches incorporated to mitigate bottlenecks.
\subsubsection{Knowledge Integration Module} Central to ThinkTank's ability to provide informed and contextually relevant outputs, this module handles all RAG-Enhanced Knowledge Integration. It enables agents to utilize information far beyond their initial training data by facilitating queries to external knowledge bases, real-time data sources, and internal organizational knowledge repositories. This grounds the agent's responses in factual data. The system supports both preset expert agents with broad domain-specific knowledge and custom agents that can incorporate proprietary organizational knowledge via user-uploaded documents, addressing the need for LLMs to utilize domain-specific information effectively while balancing utility and privacy. Underlying this RAG capability are sophisticated knowledge storage solutions like vector stores and, potentially knowledge graphs, with adaptive filtering mechanisms dynamically adjusting relevance thresholds to ensure agents receive the most pertinent information.

\subsubsection{Collaboration Framework} This component implements the sophisticated meeting structures and coordination mechanisms that define ThinkTank's collaborative process. It houses the logic for the multi-meeting, multi-round system, enabling user-defined meeting topics and iterative discussions involving the agents instantiated by the Orchestration Layer (Coordinator, Domain Experts, Critical Thinker). The framework manages how these agents interact within these defined structures, ensuring that collaborations are focused, productive, and aligned with the user's objectives. This structured approach to collaboration is vital for enhancing the overall quality and relevance of the system's output, mirroring effective human teamwork and addressing the need for dynamic teaming and conversation flow control in multi-agent systems.

\subsubsection{Local LLM Interface} A critical component for ensuring data privacy, security, and operational autonomy is the Local LLM Interface. This module provides standardized and secure access to locally deployed powerful large language models (e.g., Llama3.1 via Ollama) directly within an organization's own infrastructure. By abstracting the specifics of LLM interaction, it allows various agents and system components to leverage powerful language processing capabilities without direct exposure to external cloud services. This local-first approach is fundamental to ThinkTank's value proposition, minimizing the risks associated with transmitting sensitive information and addressing significant privacy concerns like data leakage.

\subsection{Agent Role Generalization Framework}
\begin{figure*}[t]
    \centering
    \includegraphics[width=0.9\textwidth]{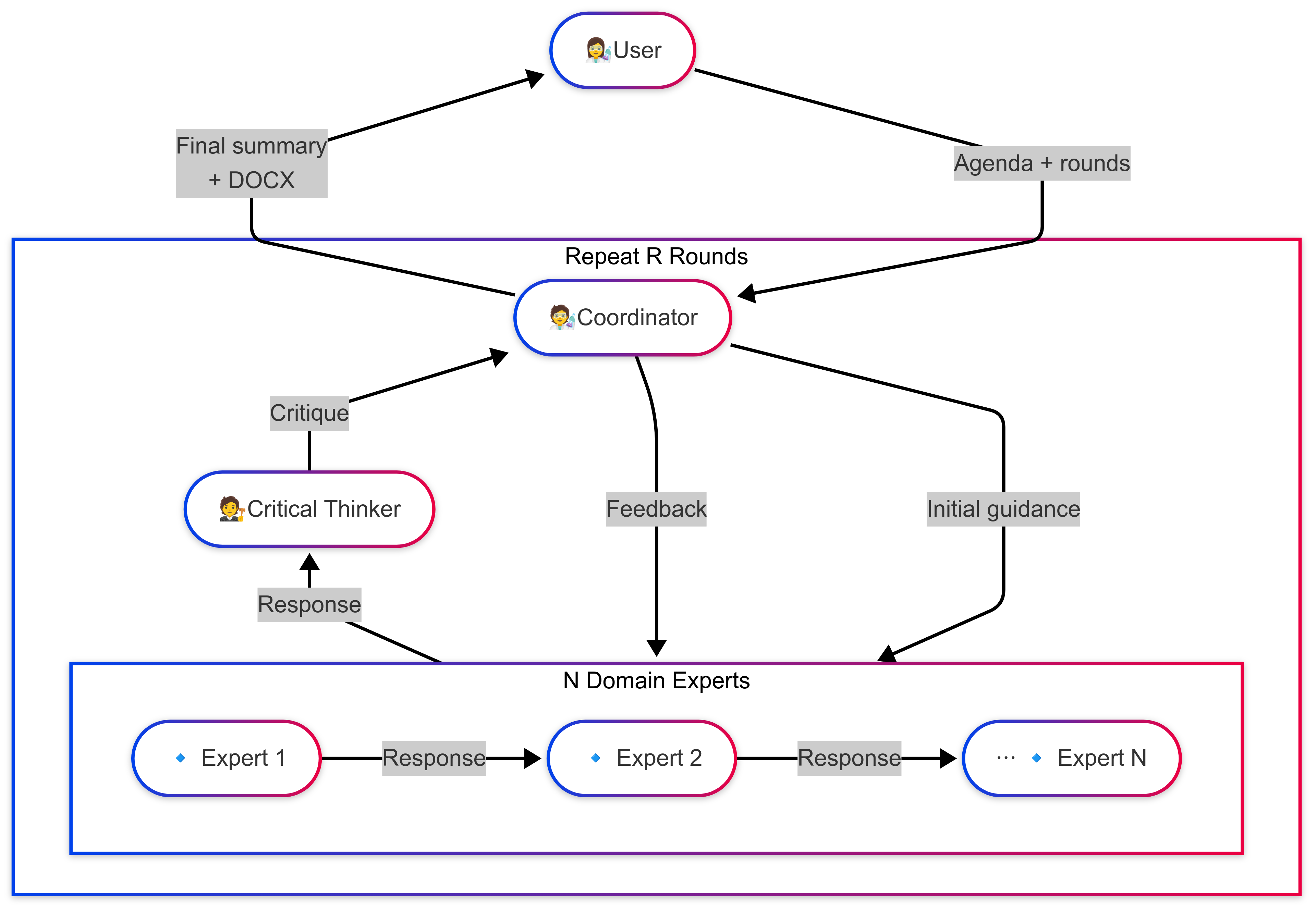}
    \caption{Schematic overview of the ThinkTank meeting workflow. The process begins with the User, who provides the agenda and specifies the number of rounds to the Coordinator. The Coordinator manages the iterative process, distributing initial guidance and feedback to a panel of $N$ Domain Experts. Each expert submits a response, which is then passed sequentially among the experts for further input. The Critical Thinker receives these responses, provides critiques, and sends feedback back to the Coordinator. The process is repeated for $R$ rounds, allowing for iterative refinement and consensus-building. At the conclusion of the process, the Coordinator compiles the final summary and documentation, which is delivered to the User.}
    \label{fig:meeting}
    \vspace{-3mm}
\end{figure*}
To achieve versatile collaborative intelligence, ThinkTank introduces an Agent Role Generalization Framework based on scientific collaboration methodologies (Fig.\ref{fig:arc}). This framework systematically adapts specialized agent roles, initially often tailored for narrow domains, into more universal archetypes capable of functioning across diverse problem-solving contexts. The core idea is to abstract the fundamental functions of these roles from their domain-specific instantiations, drawing inspiration from the proven success of well-defined roles within human scientific collaboration methodologies. 

In scientific endeavors, teams often comprise individuals with distinct responsibilities: coordinators guiding the project, specialists contributing deep domain knowledge, and critical reviewers ensuring the validity of findings. Our framework translates these collaborative dynamics into the AI agent ecosystem through functional abstraction and domain adaptation, building upon established multi-agent collaboration mechanisms.

The framework transforms three core agent types, reflecting the effectiveness of specialized yet coordinated roles observed in both scientific teams and distributed AI problem-solving scenarios:
\subsubsection{Generalized Coordinator} In scientific collaborations, a project lead or principal investigator is crucial for steering the research, synthesizing diverse contributions, resolving conflicts, and guiding decision-making. The ThinkTank Generalized Coordinator agent embodies these functions. It retains the core responsibilities of synthesis, conflict resolution, and guiding the decision-making process but expands its operational scope beyond purely scientific contexts to any defined project or complex goal. Effective coordination in multi-agent systems necessitates managing interactions between agents while ensuring system-wide coherence and maintaining alignment with overarching objectives. This agent is responsible for orchestrating the inputs from various specialized agents and producing coherent, integrated outputs, much like a scientific team leader integrates findings from different sub-teams or experiments.

\subsubsection{Critical Thinker Agent} Scientific progress relies heavily on rigorous scrutiny, peer review, and the critical evaluation of hypotheses and methodologies. The Critical Thinker Agent in ThinkTank adapts this principle of scientific rigor into a generalized role focused on logical rigor and quality assurance. It examines arguments, plans, and outputs for fallacies, unstated assumptions, potential biases, and implementation risks, mirroring how researchers critically evaluate each other's work. Multi-agent systems benefit significantly from robust reasoning, often achieved through iterative debates or discussions among agents to refine intermediate results and enhance the quality of final outcomes. While the specific evaluation criteria applied by the Critical Thinker Agent can adjust dynamically based on the domain requirements of the task at hand, it consistently maintains high analytical standards, ensuring the integrity and soundness of the collaborative output.

\subsubsection{Domain Expert Agents} Scientific collaborations bring together experts from various disciplines, each contributing deep knowledge and specialized skills (e.g., a biologist, a statistician, a chemist working on a multifaceted health problem). The Domain Expert Agents in ThinkTank are designed to be dynamically instantiated specialists. They retain their on-demand creation capabilities but can now draw upon an expanding array of expertise domains, moving beyond a single predefined area. In the ThinkTank framework, the system analyzes project descriptions or problem statements to identify the necessary areas of expertise. It then instantiates appropriate Domain Expert agents equipped with domain-specific knowledge, methodologies, and characteristic contribution styles, ready to contribute their specialized insights to the collective effort.

\subsection{Enhancing Domain Expert Agents with RAG}
The Domain Expert Agents in ThinkTank are designed as dynamically instantiated specialists, capable of drawing upon an expanding array of expertise domains. To significantly enhance their capabilities and ensure their knowledge remains relevant and deep, these agents are augmented with advanced knowledge integration and management techniques:

\subsubsection{RAG for Grounded Expertise} Domain Expert Agents leverage RAG to access and incorporate vast amounts of information, ensuring their outputs are grounded in verifiable knowledge. This is crucial as "closed-book" LLMs often struggle with domain-specific questions, highlighting the need for RAG to solve expert problems \cite{tran2025multi}. RAG helps reduce hallucinations and improves the quality of generation, especially in expert systems where inherent LLM knowledge may be insufficient \cite{bechard2024reducing}, \cite{song-etal-2024-rag}. Frameworks like DuetRAG \cite{jiao2024duetrag} demonstrate how integrating domain fine-tuning with RAG can improve knowledge retrieval quality and thereby enhance generation quality. To support RAG, Domain Expert Agents utilize robust knowledge bases structured as vector stores for efficient similarity search and knowledge graphs (KGs) for capturing complex relationships and structured domain knowledge. The integration of KGs with RAG, as seen in frameworks like RAG-KG-IL \cite{yu2025rag} and KAG \cite{liang2025kag}, provides structured domain knowledge for improved consistency, depth of understanding, and reasoning performance. This combination allows agents to not only retrieve relevant text snippets but also understand the semantic connections within the knowledge.

\subsubsection{User-Driven Knowledge Enhancement} Recognizing that expertise is often contained in proprietary or newly published documents, the ThinkTank framework allows users to upload relevant documents, research papers, and other materials directly into the system. Tool-kits like LangChain \cite{langchain} can parse various document formats into structured representations, making them accessible to the agents. This feature enables organizations to tailor the expertise of Domain Expert Agents to their specific needs and datasets, effectively creating highly specialized \textit{virtual experts}. To prevent knowledge obsolescence and ensure that Domain Expert Agents remain current with the latest developments in their respective fields, they are designed with continual or lifelong learning capabilities. This involves mechanisms for incrementally updating their knowledge bases without requiring complete retraining, thereby adapting to new information and evolving insights \cite{huang2025onpremises}. Techniques for lifelong learning in LLM-based agents focus on enabling continuous adaptation, mitigating catastrophic forgetting, and improving long-term performance by evolving their memory modules to store and retrieve updated knowledge. This ensures that the agents can learn from new data and interactions over time, maintaining their relevance and accuracy.

\subsection{Orchestrating Collaboration through Meetings}
Effective collaboration, whether human or artificial, benefits immensely from structured interaction. Just as scientific progress is often driven by well-organized meetings, workshops, and review sessions, the ThinkTank framework adapts proven scientific meeting formats to orchestrate complex problem-solving among AI agents across diverse productivity applications (Fig. \ref{fig:meeting}). This systematic adaptation is grounded in research demonstrating that structured collaboration frameworks significantly enhance performance and outcomes in various domains, including multi-agent systems. ThinkTank employs a flexible yet structured meeting system to facilitate focused and iterative collaboration among AI agents, all initiated and guided by user-defined objectives. This system is designed to break down complex projects or problems into manageable segments, allowing for deep exploration and refinement of ideas, strategies, and solutions. 

The entire collaborative process within ThinkTank begins when a user provides a description of the project or problem. This initial input serves as the foundational context for all subsequent agent interactions and meeting orchestrations. Based on this project, the user can then define and initiate a series of distinct meetings, each tailored to address specific aspects or subtopics of the larger challenge.

ThinkTank's meeting architecture is characterized by three key layers of organization:
\subsubsection{Multiple Meetings per Project} A single project or complex problem can be addressed through a sequence of different meetings. Each meeting is assigned a central topic or objective by the user, allowing for a focused discussion on a particular facet of the project (e.g., "Initial Brainstorming for Marketing Strategy," "Technical Feasibility Analysis of Proposed Feature X," "Risk Assessment of Option Y"). This modular approach enables users to systematically guide the AI agent collective through different stages of problem solving or project development, ensuring that each critical area receives dedicated attention.
\subsubsection{Multiple Rounds within Each Meeting} Once a meeting is initiated with its defined central topic, it proceeds through one or more rounds of interaction among the participating agents (e.g., Coordinator, Domain Experts, Critical Thinker). All rounds within a single meeting remain focused on that meeting's central topic. The user has the ability to specify the number of rounds a particular meeting should run. This allows for control over the depth of discussion and iteration on a specific topic. For instance, a simple information-gathering meeting might require only one or two rounds, while a complex strategy refinement meeting might benefit from several iterative rounds.
\subsubsection{Warm-up Meeting} These meetings act as a preparatory phase where a user initiates a focused session with a single domain expert agent prior to integrating them into a broader multi-agent collaboration. These one-on-one sessions serve as an opportunity for the expert to read through their domain-specific knowledge base using a RAG and build an internal memory of relevant concepts, terminology, and context. This process enables the expert to match their reasoning and knowledge level with that of other agents, thereby promoting more coherent, informed, and productive contributions in subsequent multi-agent meetings.

The \textit{Generalized Coordinator} agent plays a pivotal role in ensuring continuity and progressive refinement within each meeting: 
(1) Round Facilitation: During each round, the Coordinator guides the interaction among the agents, ensuring they contribute relevantly to the meeting's central topic. 
(2) Outcome Summarization: At the conclusion of each round, the Coordinator is responsible for synthesizing and summarizing the key discussion points, decisions, criticisms or outputs generated by the agent collective during that round. 
(3) Information Carry-Over: Crucially, the Coordinator carries forward this summary and follow-up questions to the subsequent round of the same meeting. This summarized context serves as the starting point or refined input for the next iteration, allowing agents to build upon previous conclusions, address unresolved issues, or delve deeper into specific aspects of the topic.

\subsection{Memory Management with Agno}
Memory management strategies leverage model sharing across agents to minimize resource requirements while maintaining response quality. Integrating Agno \cite{agno} into ThinkTank significantly enhances multi-agent orchestration, flexibility, and performance. Agno’s plug-and-play memory modules allow ThinkTank agents to track both short-term conversational context and long-term project knowledge, supporting deep contextual awareness and persistent learning across sessions. With Agno, we are able to address the challenge of managing various types of memory that serve different objectives in multi-agent systems \cite{han2024llm}.

\section{Preliminary Results}
\begin{figure*}[t]
    \centering
    \includegraphics[width=1.0\textwidth]{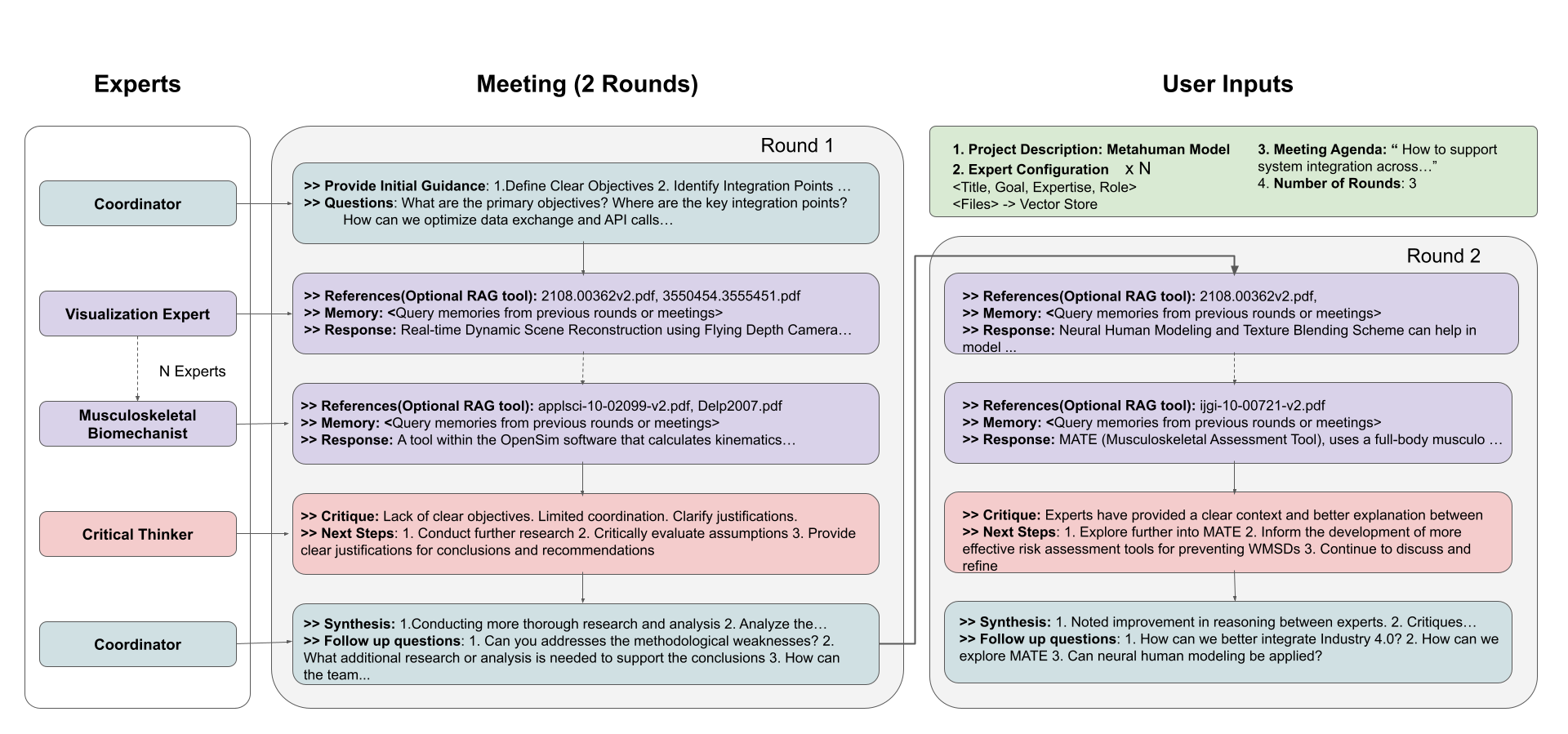}
    \caption{An example of a project meeting on the Metahuman model in the ThinkTank system. The process unfolds across two rounds of collaborative meetings, with each expert contributing domain-specific insights while referencing technical documents through optional RAG tools. Round 1 focuses on defining objectives and integration points, while Round 2 advances to a more in-depth analysis of the project of interest, which was pointed out by the Critic in Round 1. The workflow automatically incorporates user inputs, specifying project parameters, maintains memory across rounds, and concludes each phase with synthesis and follow-up questions to ensure progressive refinement of collaborative expertise.}
    \label{fig:diagram}
    \vspace{-3mm}
\end{figure*}

To validate the effectiveness of the ThinkTank framework, we conducted a demonstration using a complex technical project focused on Metahuman model development (Fig. \ref{fig:diagram}). This example illustrates how the system orchestrates multi-agent collaboration through structured meetings, leveraging specialized expertise and iterative refinement to address sophisticated technical challenges.

\subsubsection{Metahuman Model Project Demonstration}
The demonstration project involved analyzing and developing strategies for implementing Metahuman models, a complex undertaking requiring expertise across multiple technical domains. The ThinkTank system assembled a team of Domain Expert agents with relevant specializations, along with the Generalized Coordinator and Critical Thinker agents, based on the project description, expert configuration, and supporting documents provided by the user.

\subsubsection{Meeting Structure and Process Flow}
The collaborative process unfolded across two structured rounds of meetings, each building upon the previous round's outcomes while maintaining focus on the central project objectives. This multi-round approach enabled progressive deepening of analysis and refinement of solutions, mirroring the iterative nature of effective human collaboration.

\subsubsection{Round 1: Foundation and Integration Analysis}
The initial round focused on establishing clear project objectives and identifying critical integration points for the Metahuman model implementation. During this phase, each Domain Expert agent contributed specialized insights from their respective fields—ranging from computer graphics and machine learning to user experience design and performance optimization. The agents leveraged optional RAG tools to reference relevant technical documents, ensuring their contributions were grounded in current best practices and technical specifications. The Generalized Coordinator facilitated the discussion, ensuring all perspectives were heard and synthesized, while the Critical Thinker agent evaluated the proposed approaches for potential risks and logical consistency.

\subsubsection{Round 2: Deep Technical Analysis}
Building on the foundation established in Round 1, the second round advanced to more detailed technical analysis. Notably, the Critical Thinker agent had identified specific areas requiring deeper investigation during the first round, which became the primary focus of Round 2. This demonstrates the system's ability to adaptively respond to emerging insights and redirect collaborative efforts toward the most critical aspects of the project. The agents engaged in more sophisticated technical discussions, exploring implementation challenges, performance trade-offs, and integration strategies with greater depth and specificity.

\subsubsection{System Capabilities Demonstrated}
The demonstration highlighted several key capabilities of the ThinkTank framework:

\textbf{Automatic Workflow Integration:} The system seamlessly incorporated user inputs and project parameters without requiring manual intervention, demonstrating the effectiveness of the Dynamic Agent Creation Engine in parsing project requirements and assembling appropriate expertise.

\textbf{Cross-Round Memory Management:} Information and insights from Round 1 were automatically carried forward to Round 2, with the Coordinator agent maintaining context and building upon previous conclusions. This memory persistence ensured continuity and prevented redundant discussions.

\textbf{Progressive Refinement:} Each round concluded with synthesis of key findings and the generation of targeted follow-up questions, ensuring that subsequent rounds would address the most important unresolved issues and advance the project meaningfully.

\textbf{RAG-Enhanced Expertise:} Domain Expert agents effectively utilized RAG tools to access and reference technical documentation, demonstrating how the Knowledge Integration Module enhances agent capabilities beyond their base training data.

The successful completion of this demonstration showcases the feasibility of ThinkTank's framework in a non-trivial technical setting, validating the core premise that systematic generalization of scientific collaboration methodologies can create effective AI-driven platforms capable of tackling complex, multi-domain challenges through structured, iterative engagement. While this represents a proof of concept implementation, the demonstration establishes a solid foundation for future development and refinement of the framework, demonstrating its potential to address sophisticated real-world problems through coordinated multi-agent collaboration.

\section{Discussion}
As a lack of standardized protocols can lead to scenarios where multiple agents speak different languages, successful deployment requires addressing organizational change management and user adoption considerations, as well as technical challenges including scalability, interoperability, and complex agent interactions \cite{han2024llm}. 
The most intricate challenge lies in managing complex interactions between agents, as the web of relationships and dependencies grows exponentially with system sophistication. Coordinating actions, resolving conflicts, and maintaining system-wide coherence requires technical prowess and a deep understanding of how autonomous entities can work together toward common goals.

Advanced agent collaboration models incorporating emerging AI technologies provide promising directions for scaling capabilities while maintaining computational efficiency. We envision that the integration with computer vision, multi modal processing, and advanced coordination mechanisms will expand collaborative capabilities beyond text-based interactions \cite{tran2025multi}.
As presented by Schmidgall \& Moor in their paper, autonomous agents may play a role in designing future AI systems alongside humans, with frameworks like AgentRxiv enabling agents to collaborate toward research goals and accelerate discovery \cite{schmidgall2025agentrxiv}. The future of multi-agent systems involves advancements in coordination algorithms, machine learning integration, and scalability solutions that will expand AI-driven problem-solving capabilities.

\section{Conclusion \& Future Work}
This paper presents a comprehensive framework for transforming specialized AI agent systems into universal collaborative intelligence platforms. The proposed approach successfully generalizes proven scientific collaboration methodologies while maintaining core benefits of multi-agent collaboration, local deployment, and expertise creation.
The framework addresses critical organizational requirements for data privacy, cost control, and system customization while providing sophisticated collaborative capabilities across diverse domains. 

One of the future research directions includes a systematic evaluation of MAS, advanced agent collaboration models, integration with emerging AI technologies, and optimization of organizational adoption strategies. The foundation provided by this generalization approach creates a platform for continued innovation in collaborative intelligence that adapts to emerging organizational needs while maintaining structured, multi-perspective problem solving enhanced by AI.
Organizations seeking to leverage AI for complex problem solving in diverse domains will find that this framework provides a practical, scalable solution that balances technological sophistication with operational requirements and strategic objectives while ensuring privacy and security through local deployment.

Another research direction involves expanding the ThinkTank framework to incorporate a decentralized open marketplace where users can share, discover, and monetize their specialized vector stores and Domain Expert agents. This marketplace would enable organizations to leverage collective intelligence by accessing curated expertise from other users while maintaining data privacy and security through controlled sharing mechanisms. Such a platform could facilitate the creation of a collaborative ecosystem where specialized knowledge bases and expert agents become reusable assets, potentially incorporating blockchain-based monetization models and reputation systems to ensure the quality and trustworthiness of shared resources. This expansion would transform ThinkTank from an isolated organizational tool into a broader collaborative intelligence network, enabling cross-organizational knowledge sharing while preserving the framework's core principles of local deployment and data sovereignty.

\bibliographystyle{abbrv}
\bibliography{references.bib}
\end{document}